# Synthesis and acid-resistance of Maya blue pigment

M. Sánchez del Río[1], P. Martinetto [2], C. Reyes-Valerio[3],

E. Dooryhée[2], and M. Suárez [4]

[1] European Synchrotron Radiation Facility, BP220, F-38043 Grenoble Cedex, France

[2] Laboratoire de Cristallographie, CNRS, BP166 F-30842 Grenoble, France

[3] Instituto Nacional de Antropología e Historia, México DF, Mexico

[4] Universidad de Salamanca, Departamento de Geología, E-37008 Salamanca, Spain

## Abstract

Maya blue is an organo-clay artificial pigment composed of indigo and palygorskite. It was invented and frequently used in Mesoamerica in ancient times (VIII-XVI cent.). We analyse in this paper one of the characteristics of Maya blue that attracted the attention of scientists since its re-discovering in 1931: its high stability against chemical aggressions (acids, alkalis, solvents, etc), and biodegradation, that has permitted the survival of many artworks during centuries in hostile environments, like the tropical forest. We reproduced the different methods proposed to produce a synthetic pigment with the characteristics of the ancient Maya blue. The stability of the pigments produced using either palygorskite or sepiolite has been analysed by performing acid attacks of different intensity. The results are analysed in terms of pigment decolouration and destruction of the clay lattice, evidenced by X-ray diffraction. Palygorskite pigments are much more resistant than sepiolite pigments. It is shown that indigo does not protect the clay lattice against acid aggression. We show that Maya blue is an extremely resistant pigment, but it can be destroyed with very intense acid treatment under reflux.

## Keywords

# 1 Introduction

Maya blue is an important pigment in Mesoamerican history. It is probably original from the Maya area in America. It was widely used in murals, pottery and sculptures in a wide region of Mesoamerica in the pre-Hispanic time (from the VIII century) and during the colonization until 1580.

The Maya blue pigment was discovered in 1931 by Merwin (Merwin 1931) when analyzing the pigments of a mural paint in the Chichén-Itzá archaeological site (Yucatán, Mexico). He thought it was an inorganic natural pigment because of its high stability to acids. Gettens named this pigment Maya blue (Gettens & Stout 1942). It was established in the 1960's (Cabrera Garrido 1969; Gettens 1962; Kleber, Masschelein-Kleiner & Thissen 1967; Van Olphen 1966) that Maya Blue consisted of two ingredients: indigo, which was produced in Central America from the leaves of the añil plant (*Indigofera suffruticosa*), and palygorskite, a clay mineral of fibrous nature found in Yucatán Peninsula, known by the Mayas (Arnold & Bohor 1975, 1977). It is reported in literature that some archaeological samples of Maya blue contain another fibrous clay, sepiolite, in addition to (or instead of) palygorskite (Gettens 1962; Sheppard & Gottlieb 1962). Van Olphen (Van Olphen 1966) obtained synthetic Maya blue resistant to acids using palygorskite and sepiolite, both clays with fibrous structure, but non resistant pigments were obtained using other clays with laminar structure (e. g. kaolinite, nontronite, etc.). However, the detailed nature of the chemical bond between the clay and the organic molecule is still not known. Two hypotheses were proposed in the sixties: Van Olphen suggested that the organic molecules are too large to enter into the channels of the clay fibers, hence the channels are in some way sealed in their ends by the indigo. Kleber *et al.* (Kleber, Masschelein-Kleiner & Thissen 1967), looking at the dimensions of the channels and indigo molecules, affirmed that the partial (or even deep) penetration of the indigo inside the channels cannot be excluded. They suggested that the irreversible fixation of indigo is associated with the loss of zeolitic water (hygroscopic water is lost below 100 ºC, zeolitic water at approximately 150 ºC and coordinated and structural water at 375-425 ºC). Recently, Hubbard *et al.* (Hubbard et al. 2003) performed thermal and textural analyses, combined with multinuclear magnetic resonance on synthetic Maya blue. They proposed a model where the indigo molecules covering the openings of the clay channels are anchored by hydrogen bonds of their carbonyl and amino groups to silanol groups bordering the micropores. Other authors (Chiari, Giustetto & Ricchiardi 2003; Fois, Gamba & Tilocca 2003) suggest that molecules do enter into the channels replacing water and occupying stable positions. A third hypothesis states that iron impurities coming from the *Indigofera suffruticosa* plant that supplies the indigo, seal the channels and are also responsible of the blue colour (Jose-Yacaman et al. 1996). It is very improbable in our opinion, and it has also been refused by other authors (Chiari, Giustetto & Ricchiardi 2003; Fois, Gamba & Tilocca 2003; Hubbard et al. 2003; Reinen, Köhl & Müller 2004).

The scientific analysis of objects containing this pigment could give the necessary information to identify a possible common origin of the pigments or the raw materials, deduce possible routes of trade and distribution, and establish the time and geographical evolution of Maya blue. This would answer some of the questions about Maya blue that Mesoamerican archaeologists and anthropologists are trying to solve.

However, for the interpretation of the results of the analysis of archaeological samples, it is often necessary to compare them with well-characterised references produced under controlled conditions in a laboratory. The goal of this work is to reproduce and compare the different methods proposed to make Maya blue, and perform a detailed analysis of one of the most important features of Maya blue, its chemical stability against acid aggressions.

# 2 Materials and experimental methods

## *2.1 Materials*

The materials used in this work come from different sources. *Añil* leaves (*Indigofera suffruticosa*) were collected and cultivated by Reyes-Valerio in Mexico. Synthetic indigo and indoxylacetate have been purchased at Sigma-Aldrich. Three palygorskites have been used, from Mexico, Senegal and USA. The Mexican palygorskite (almost free of impurities) comes from a mine close to Ticul (Yucatan). Palygorskite from Senegal was provided by the TOLSA company. Its purity is about 75% (opale impurities rise to 20%). Palygorskite from USA, obtained though the Source Clays Repository of the Clay Minerals Society, is from Attapulgus (Georgia) and has less than 10% of impurities (quartz and smectite). The montmorillonite, a planar clay, comes from the latter supplier. Sepiolite is from Yunclillos (Toledo, Spain), it has a purity of 90-95%, with impurities of quartz, feldspars and calcite, and it was also supplied by TOLSA. It also delivered a purified and micronized industrial product (Pangel S-9) made from this sepiolite.

Palygorskite (formerly attapulgite, name still in use for industrial applications) is a clay mineral. Together with sepiolite, it forms the group of fibrous clay minerals. The structure of this mineral proposed by (Bradley 1940) contains ribbons of 2:1 phyllosilicate structure linked by periodical inversion of the apical oxygen of the continuous tetrahedral sheet. Therefore, the octahedral sheet is discontinuous (Fig. 1). The theoretical formula is $Si_8 O_{20} Al_2 Mg_2 (OH)_2 (H_2O)_4\ 4(H_2O)$; as can be observed in Fig. 1, there are three different types of molecules of water in this formula: hydroxyl groups, water bonded to octahedral cations and adsorbed water (named zeolitic water). Zeolitic waters occupy channels parallel to the fibre axis.

Palygorskite from Sacalum in Yucatan Peninsula has served as a major source of this clay for Maya Blue for over 800 years. The clay occurs as a lens approximately one meter thick and its origin is marine sedimentation (Isphording 1984). It occasionally contains sepiolite as impurities.

## *2.2 Acid tests*

For the acid treatments, we have used acids from commercial suppliers of chemical products. Acids were used in their original concentration (which we call "concentrated" acids hereafter), which are about 70% for $HNO_3$ (i.e. about 15 mol.l$^{-1}$) and 37% for HCl (12 mol.l$^{-1}$). With hot acids, the heating is done *under reflux*, that enables a mixture containing volatile materials to be heated for a long time without loss of solvent. It consists in a flask with a water-refrigerated column (condenser) on top of it, which condenses the vapour and returns it to the flask. This method is widely used for acid treatments in Chemistry, and has rarely been applied to the study of the Maya blue. In reflux experiments the "concentrated" acids have been diluted with the same volume of water, giving therefore concentrations of about 7 mol.l$^{-1}$ for HNO$_3$ and 6 mol.l$^{-1}$ for HCl.

To discuss our experimental results, it is useful to classify the acid tests (with acid in excess) in different types:

1) Samples are first tested in concentrated HNO$_3$ or HCl during 5-15 minutes, at room temperature in a watch glass.

2) A more aggressive nitric acid test was applied for 1.5 hours, with concentrated acid at room temperature.

3) Some samples were left in concentrated acid at room temperature during two to four days.

4) Reflux experiments. A solution of concentrated acid and diluted in water in a 1:1 proportion is heated at a temperature close to ebullition (90 ºC) during five hours or 30 hours in some cases.

## 2.3 Powder x-ray diffraction

The samples were analysed by X-ray powder diffraction, before the acid attacks, in order to study the changes in the crystalline structure.

The X-ray diffraction patterns have been collected using a Bruker AXS D8 diffractometer equipped with a Cu Kα x-ray tube, a Ge(111) incident monochromator, and a solid state detector (MXP-D1, Moxtec-SEPH), which filters the fluorescence x-rays and records only the x-rays scattered elastically. A small quantity of the sample was placed onto a kapton foil and mounted in transmission mode. Results of different measurements are normalized in counts/sec in order to be directly compared. The mass of the sample also contributes to the diffraction profile as a scale factor, but it has not been controlled.

Acid treatments are usually done on clay minerals, including sepiolite and palygorskite, to increase their chemical activity, because they modify the clay surface area and/or the number of active sites (Suárez Barrios et al. 1995). The acid dissolves the Mg, Al and other cations in octahedral sites. As a result of this dissolution, contiguous tetrahedral sites with Si in the center are altered, loosing their crystal structure, and forming amorphous silica. This effect is evidenced by X-ray diffraction. Characteristic peaks of the clay disappear, and a broad halo at $2\theta \sim 22°$ (for Cu K$\alpha$ radiation) is produced by the Si-Si correlation function of the amorphous silica (Warren 1990). The amount of destruction of the clay depends on the type of acid, acid concentration, temperature, and exposure time. When the total destruction of the clay is achieved, the diffractogram presents no peaks, and only the amorphous silica halo can be observed.

One of the historical problems to identify the ingredients of Maya blue has been the difficulty to detect the indigo in the pigment. X-ray diffraction patterns of the pigment are substantially the same as that of the clay used for the preparation. It has been reported (Kleber, Masschelein-Kleiner & Thissen 1967) the difficulty to identify peaks corresponding to indigo in the pigments prepared with indigo concentration less than 5%. We believe that it is not possible to detect indigo peaks in the XRD pattern of the pigment not only because of the small concentration of indigo in the Maya blue (usually 1% or less) but also because the crystalline structure found in synthetic indigo powder is lost when combining with the clay. The XRD technique could be useful in identifying pure indigo if sufficient material was available, but this is rare with pigments and colorants on objects or paintings.

# 3 Methods for preparing the Maya blue pigment

Several methods have been proposed for producing artificial pigments with characteristics (colour and stability) similar to those of the Maya blue found in archaeological samples. Most methods were proposed with the aim of obtaining a pigment whose characteristics are compatible with those of the ancient Maya blue, without questioning whether Mayas would have been able to prepare the pigment in the same way. Other preparation methods, as the one suggested by Reyes-Valerio (1993), pay special attention to the fact that the ingredients and the preparation methods should be compatible with the resources and technology accessible to the Mayas, more than one thousand years ago. Van Olphen (1966) suggested three methods for producing the Maya blue: from a solution of indoxylacetate, by vat-dyeing and powder mixture, and by mixing up the powdered components. However, very few details of the preparation were given in the paper. Kleber *et al.* (1967) used the procedure of mixing indigo and clay powders to produce the pigment, and they described the indigo concentration, heating process and stability test results. The traditional method described by Reyes-Valerio (1993) is summarized and the slight modifications used to reproduce this recipe are given. We have not considered other recipes proposed by Littman (1982) because they do not apply a heating treatment to the dried pigment. This has been demonstrated to be essential for the stabilization of the pigment in all other preparations found in literature. However, it is interesting that Littmann used HCl as a reducing agent, and it is the only reported experiment, in addition to those of Reyes-Valerio, that used añil leaves as a source of indigo.

### *3.1 Solution of indoxylacetate*

This preparation method was proposed by Van Olphen (1966): *"When a slightly alkaline solution of synthetic indoxylacetate contacts attapulgite, the clay becomes blue; ..."*.

We mixed about one gram of palygorskite with a few milligrams of indoxylacetate in 10 ml of distilled water and some drops of a concentrated solution of NaOH to obtain a pH of 8. We heated to 110 °C to evaporate the water. When the sediment was dried, we added 10 ml water and heated again. This operation was performed four more times. After that, the residual powder presented a blue tonality, but it was decoloured once in contact with nitric acid. It became resistant to concentrated nitric acid during a few minutes after being heated in an oven during 5 hours at 190 °C.

### *3.2 Vat-dyeing*

This preparation technique was also proposed by Van Olphen (Van Olphen 1966). It is used worldwide to dye clothes (cotton, wool, silk, etc.) with indigo. In particular, all blue jeans are dyed using this technique. Indigo is not soluble in water, so it has to be reduced to a soluble leuco-base form (which is yellow), then contact the palygorskite, dry it and oxygenate with air. During the oxygenation the clay becomes blue.

We prepared a mixture of 0.1 g of synthetic indigo with a few drops of ethanol. Then it was digested in about 5 ml of distilled water, and 2 ml of 2M NaOH was then added. After agitation, we prepared another solution with 20 ml water, 0.3 g of sodium hydrosulfite. We mixed both solutions and heated the new solution at 50 °C for 40 minutes. After this time, the solution appeared yellow, however, a blue layer could be observed on the surface (as a result of oxygenation with air). Then one gram of palygorskite powder and 40 ml water were added. The solution is agitated with a magnetic stirrer while being heated at 50 °C. After one hour we filtered the sediment and put it on a glass. The sediment looked yellow but became blue in seconds. After one night we collected the dried powder. We observed that it was decoloured once in contact with nitric acid, but, as before, it became resistant after being heated in an oven during 5 hours at 190 °C. This result disagrees with what was reported by (Littmann 1982), who obtained stable pigments by only heating the solution and without heating the dried pigment.

### *3.3 Using Indigofera leaves*

This method uses natural leaves of the *Indigofera suffruticosa* plant. This plant was known in pre-Hispanic times. The plant is locally known by its Spanish (añil) or Nahuatl (xiuquilitl) names. This method was proposed and fully described in (Reyes-Valerio 1993). This preparation method aims at reproducing the procedure that Mayas might have used to prepare the pigment. It only uses materials that were available to the Mayas (water, leaves and palygorskite powder) and only require simple manipulations. The original recipe was partially described by Francisco Hernández in the XVI century (Hernández 1959). According to him, the Indians obtained an original dark blue product formed by the extract of leaves of añil (leuco-indigo) adsorbed in clays. However, Hernández did not mention that the heating process is necessary to make the clays resistant to chemical aggressions. Reyes-Valerio succeeded in making synthetic acid resistant Maya blue using sacalum (palygorskite) and *añil* leaves. Several detailed formulas to prepare the pigment are described in detail in his book (Reyes-Valerio 1993). He followed these steps:

1. Fermentation or maceration. In a recipient of 0.5 l the *añil* leaves (fresh or dried) are placed in a proportion of 3-5 g for each 100 ml of distilled water. Then 1.0-1.5 g of palygorskite (perhaps mixed with other clays) is added. The amount of clays must not overpass the limit of 1.5 g per 5 g of leaves in 100 ml of water. These ingredients are stirred frequently, by rotating the recipient or with a glass stick. The maceration time is variable, no longer than 24 h for fresh leaves (15-18 h for dried leaves). As a sign of the correctness of the process, an iridescent layer is formed on the liquid surface. The liquid becomes turbid with a greenish or light blue tonality.

2. Removal of leaves, agitation and oxygenation. The leaves are removed using a mesh. The liquid is then oxygenated (during about 10 minutes) by passing from one recipient to another, and repeating the process, or by using a stirrer (or a combination of these two methods). During this step, a green-blue foam is formed at the surface. After about 10-15 minutes of agitation, the foam loses its colour. The process is then stopped, and the mixture is let to rest during 30 minutes. Then the dyed clay particles should be deposited at the bottom of the recipient.

3. Filtering. With a paper filter (like the number 1 of Whatmann) the mixture is filtered. The filtered liquid (of a yellowish colour) is wasted.

4. Dried and heating process. The filter paper with the sediment is placed in a laboratory heater. The temperature should not be higher than 90-100 °C. The clays change colour from dark blue to other more or less lighter blues. After 20-30 minutes, they become turquoise. The process ends here. The stability of the resulting pigment can be tested with nitric acid or any other acid or base or aqua regia.

In our preparation we used *añil* leaves and Mexican palygorskite (sacalum) and followed these instructions. The first trial did not produce a blue pigment but rather a greenish one. However, it was acid-resistant after heated. The lack of success was certainly due to insufficient maceration and fermentation of the leaves.

We slightly modified the recipe and managed to obtain a good-looking stable pigment: We mixed 0.25 g of palygorskite, 0.6 g of dried *añil* leaves, and 25 ml of distilled water. We agitated the mixture during ten minutes using a magnetic stirrer, and then stopped the stirring and waited for 20 minutes. We repeated this operation during 8 hours (16 times). After this, we stopped stirring and let the solution macerate overnight (16 h). Then we did not observe a surface layer of foam as predicted by Reyes-Valerio (1993), but we continued the experiment. We separated the leaves from the liquid with a grid, and put the liquid in another recipient placed on the magnetic stirrer during 20 minutes to oxygenate the mixture. After Reyes-Valerio (1993), we should have seen a layer of coloured foam, but we did not observe anything. We filtered the clays and then dried overnight at room temperature on the paper filter. The resulting blue mixture was grinded in a mortar and then heated in an oven for 1.5 h at 90 °C. The resulting pigment was acid-resistant. A previous experiment using the same preparation method, but stirring during 3 hours instead of 8 gave a stable pigment but the colour was greenish.

It is worth noting that we obtained, in our first try, a pigment that looks more green than blue using the traditional method of preparation with añil leaves. This was probably due to a non-complete maceration and oxygenation of the leaves, or an insufficient amount of leaves. It is interesting that blue in Bonampak murals (Chiapas, Mexico, VIII cent.) have a very strong greenish colouration, and in many places it is a very bright green. It is therefore possible that Mayas could have used the same method both for the Maya blue and for the green pigment in Bonampak. Further studies are required from the archaeological point of view (characterising the composition of Bonampak's green) and for synthesising a green pigment using a variant of the Maya blue producing mechanism.

### *3.4 Mixture of synthetic indigo and clay powders*

This technique consists in preparing a mixture of synthetic indigo powder and the clay, and heating it for stabilization (to make it resistant to acids and chemicals). This was proposed by Van Olphen (1966) and then reused by Kebler Masschelein-Kleiner & Thissen (1967) and others. This technique is the easiest and simplest way to prepare the pigment, and allows controlling the parameters (mass of indigo and clays, etc.) with more accuracy because no solutions are involved.

We have prepared mixtures of synthetic indigo and different clays (palygorskite, sepiolite and montmorillonite). We used different concentrations of indigo: 10, 7, 5, 3 and 1%. The powders were finely grinded and mixed in a manual mortar. The resulting powder was sometimes pressed into a pellet for better manipulation and to improve the contact between the clay and indigo. The pellets or the powder were placed in an oven and heated. Our "standard" heat treatment (190 °C for 5 hours) has been decided by looking at published results (Kleber, Masschelein-Kleiner & Thissen 1967). This heating treatment is the key point for stabilization of the pigment. The temperature and time were chosen "in excess" in such a way that we are sure that the stabilization of the pigment is achieved. Stabilization can also be produced with lower temperature and time, and this is discussed later. After that, the pellets were re-grinded and sometimes washed with acetone to dissolve the indigo that was not adsorbed by the clay. This is done by adding acetone to the pigment on a filter paper, until non-coloured acetone flows.

The pure indigo powder presents a very dark blue colour, close to black. We found that, when mixed with palygorskite (white) to create the Maya blue, a concentration of 1% is enough for giving a light blue colour close to what is expected for the Maya blue. Moreover, after heating the mixture to produce the pigment, the tonality changes, and it tends to approach more a turquoise-greenish colour, as reported in literature for the Maya blue. This change of colour during heating has been observed and reported in traditional preparations (Reyes-Valerio 1993). Moreover, a recent optical spectroscopy study (Reinen, Köhl & Müller 2004) showed a shift of the blue colour to the turquoise and green when the indigo-palygorskite mixture is heated, indicating a chemical interaction. When adding more indigo to the mixture (3%), the pigment becomes more dark blue, much different from the archaeological Maya blue. In many cases it is very difficult to appreciate changes in colour, and further quantitative studies using colorimetry and optical spectroscopy should be done, comparing archaeological and new samples synthesised under different conditions. We find that palygorskite can be mixed with indigo at different concentrations: concentrations as high as 10% give stable pigments, even though the colour does not correspond to the Maya blue. We obtained different colours depending not only on the concentration of indigo, but also on the palygorskite origin (Mexico, Senegal or Attapulgus), and on the grinding of the powder (done manually in a mortar).

We observed that all pigments prepared using fibrous clays (sepiolite and palygorskites) after being heated at 190 °C for 5 h were not decoloured when treated with concentrated $HNO_3$ or HCl for several minutes. We observed an immediate decolouration for the planar clays, as expected following (Van Olphen 1966). We tested montmorillonite because this author did not test it, and it was proposed to play a fundamental role in the Maya blue (Littmann 1980) (although this idea was later discarded (Littmann 1982)). We confirmed that a pigment made with montmorillonite (after heated 5 h at 190 °C) is immediately decoloured when it contacts concentrated $HNO_3$.

Almost all preparation methods have in common a heating phase after the ingredients have been combined and dried. The strength of this thermal treatment, reported in the literature, is very variable: the temperature ranges from lower than 90-100 °C (Reyes-Valerio 1993) to 190 °C (Kleber, Masschelein-Kleiner & Thissen 1967) or even 250-300 °C (Littmann 1982), while annealing time ranges from a few minutes (Reyes-Valerio 1993) or hours to several days (Van Olphen 1966). We found that palygorskite-indigo mixtures should be heated at least at 80-90 °C (we heated during 1.5 hours) in order to be acid-resistant. This temperature agrees with the experimental methods in (Reyes-Valerio 1993) who heated to 90-100 °C during 20-30 minutes. Most of the hypothesis about the high resistance of the Maya blue assign the stabilization to the loss of zeolitic water (Chiari, Giustetto & Ricchiardi 2003; Kleber, Masschelein-Kleiner & Thissen 1967), and sometimes the elimination of some structural water (Fois, Gamba & Tilocca 2003). However, the stabilization temperature can be as low as 80-100 °C, as we observed experimentally. This temperature mostly corresponds to the elimination of hygroscopic water. Zeolitic water is lost at slightly higher temperatures (120-130 °C). Further thermal studies are required to quantify the amount of zeolitic water that could be eliminated from the palygorskite as a function of time.

## 4 Acid resistance

Almost all the authors state that the Maya blue presents an exceptional stability to chemical aggressors (acids, alkalis and organic solvents). However, with a few exceptions, the tests reported in literature (Gettens 1962; Littmann 1982; Van Olphen 1966) are not quantified in terms of concentration, temperatures, and duration of the attacks nor compared with identical tests performed independently on indigo and clays contained in the Maya blue. Moreover, the results are drawn from a visual inspection of the loss of pigment colour, and no other test hase been carried out to check if the molecular or crystallographic structures of the indigo and clay have changed. This can be the origin of many uncertainties, as it is sometimes difficult to appreciate colour changes by eye. Moreover, a chemical change is not always accompanied by a change in colour.

Our motivation is to quantify the resistance of the Maya blue pigment and its ingredients to acid attacks. We study how an acid attack alters not only the colour, but also the crystallographic structure of the synthetic Maya blue pigment. The results are compared with the alteration of the raw clay under the same conditions.

The samples used for the acid resistance studies reported in this paragraph are prepared by heating a mixture prepared using synthetic indigo and clay. This is the simplest method of preparation, and permits, from the chemical point of view, a more accurate control on the concentration of the ingredients.

## *4.1 Indigo*

Indigo ($C_{16}H_{10}N_2O_2$) is an organic colorant widely used for dyeing textiles and as a colorant for artistic pigments. It can be obtained from the leaves of some plants, the most known are *Indigofera tinctoria* from India, *Indigofera suffruticosa* from America and *Isatis tinctoria* in Europe. Nowadays it is chemically synthesised. In the pure form, indigo is presented as a dark blue powder. It is insoluble in water. It melts at about 390 ºC and sublimes undecomposed. Indigo powder is presented in form of rhombic grains giving a well-characterised diffraction pattern.

We have tested the resistance of the colour of synthetic indigo powder to different acids. It is not decoloured by concentrated hydrochloric acid, and dissolves in sulphuric acid (98%) keeping its dark blue colouration. When the dark blue indigo powder contacts concentrated nitric acid, it converts into a yellowish-red precipitate. The reason is that strong oxidizing agents, like concentrated nitric acid, decompose the indigo into isatin ($C_8H_5NO_2$). This effect is also manifested in the fading of the indigo and other organic colorants in artworks (Grosjean, Salmon & Cass 1992) as a result of atmospheric nitric acid in polluted environments. Therefore, for our resistance tests, the use of $HNO_3$ is preferred with respect to other acids (HCl, $H_2SO_4$) because it destroys and decolours immediately the indigo powder.

## *4.2 Palygorskite and sepiolite*

Both sepiolite and palygorskite are resistant to concentrated acids ($HNO_3$ and HCl) at room temperature during a few minutes. This is confirmed by the fact that the diffraction patterns of the attacked samples are identical to those of the samples before attack.

If sepiolite is immersed in concentrated $HNO_3$ during 1.5 h, the peaks in the XRD pattern start to decrease, meaning that the structure starts to decompose (Fig 2). Longer attacks with concentrated nitric acid (2-7 days) destroy completely the structure producing amorphous silica.

Palygorskite (we tested Mexican palygorskite) shows to be resistant to concentrated acids (HCl, $HNO_3$) at room temperature during 2-3 days. No change in the diffractogram was observed, contrary to what happened to sepiolite, which was completely destroyed under the same conditions. To perform attacks at higher temperature, we used reflux experiments with 6 mol.l$^{-1}$ HCl and 7 mol.l$^{-1}$ $HNO_3$ acids, at a temperature of 90 ºC during 5 hours. Fig 3 shows that the palygorskite structure is severely altered. A more intense treatment with the same concentration of $HNO_3$ at 90 ºC during 24 h hours did not destroy completely the palygorskite although it caused severe damage. Total destruction of the palygorskite is possible if the treatment is strong enough (for example 24 h at boiling temperature with 6 mol.l$^{-1}$ HCl, as suggested by Gonzalez et al. (1989)). We performed reflux with the same concentration of HCl during 30 hours at boiling temperature and the diffractogram showed no palygorskite at all (Fig. 3).

### *4.3 Sepiolite-based pigments*

Acid tests were done with pigments prepared with sepiolite from Spain (raw mineral and Pangel product) mixed with 1% and 3% of indigo.

We observed a good resistance to concentrated acids at room temperature during a few minutes for pigments prepared in the usual way (heated 5 hours at 190 ºC using 1% and 3% of indigo).

We reduced the heating time and prepared new samples heated for 1.5 hours at 90, 70 and 50 ºC, with good resistance to decolouration. We tested the non-heated mixture and we also found it resistant in the case of the pigment prepared with Pangel, but not with the pigment prepared with the mineral sepiolite, which was decoloured. These results indicate that the industrial processing of the mineral sepiolite to produce the Pangel product is the clue for the stabilization of the unheated samples. This process consists in a fine grinding in humid environment and separation by grain size, thus merely physical treatments that do not modify the aspect-ratio of the individual particles with respect to the mineral grains. We confirmed by DTA (Differential Thermal Analysis) that similar curves were given by the mineral of sepiolite and the Pangel product. Furthermore, by very fine (manually) and intensively grinding the raw mineral of sepiolite in a mortar, we also obtained a stable pigment without any heating.

The resistance to concentrated nitric acid for several minutes of unheated mixtures of sepiolite and indigo was achieved with industrially prepared sepiolite (including fine grinding) and also with the mineral sepiolite grinded very intensively in a mortar. This indicates that smaller grain size improves the formation of the indigo-sepiolite complex. Similar results of stable non-heated sepiolite-based pigments are reported in Hubbard et al. (2003), where it is confirmed by magnetic resonance that the sepiolite-indigo complex starts to form prior to the heating process. Moreover, the fact that we made pellets when prepared some samples could also favour the indigo-sepiolite interaction in the resulting pigment. It is therefore possible to stabilize sepiolite and indigo even with no heating (this effect is not true for palygorskite, see later) The grain size is an important parameter in the stabilization. This reduction of the grain size increments the external surface of the clay and makes more accessible the silanol groups at the edge of the fibre (Si-OH). Whether or not the indigo reacts with these silanol groups is an open question.

Longer acid tests (1.5 hours) on heated pigments strongly alter the colouration, and the results showed different bluish-grey colourations. However, all samples lose their blue tonality after 2-4 days in concentrated acid. In Fig 4 is plotted the diffraction patterns of sepiolite pigment after intense acid treatments (1.5 h with nitric and four days in nitric and hydrochloric acids), that decoloured completely all pigments. For the 1.5 h attack, the diffractogram shows a partial destruction of sepiolite, characterised by the reduction of the peak intensity, in a similar way as it is observed for the raw sepiolite treated during 1.5 h. It is remarkable that the complete decolouration here is not accompanied by a total destruction of the sepiolite. Longer attacks at room temperature during four days completely destroyed the sepiolite structure, and the diffractograms show only amorphous silica.

## *4.4 Palygorskite-based pigments*

The unheated mixtures of palygorskite-indigo were always immediately decoloured when immersed in nitric acid. In order to compare with the case of sepiolite pigments, where we obtained good resistance for the case of the product prepared for industrial applications, we fabricated one pigment using an industrial palygorskite produced using the same treatment applied to sepiolite (Pangel). Even in this case we found an immediate decolouration.

Heated mixtures (pigments) showed to be resistant to acids at room temperature for long times. Even if treated for four days in concentrated HCl or $HNO_3$, they were not decoloured. Moreover, the XRD profiles are identical to that of the raw palygorskite, showing complete stability of the pigment to acid attacks at room temperature.

We treated several pigments under reflux (5 h at 85-90 ºC) with acids (6 mol.l$^{-1}$ HCl and 7 mol.l$^{-1}$ $HNO_3$). All pigments were decoloured, ranging from total to partial decolouration. A pigment made with Mexican palygorskite and 1% synthetic indigo was completely decoloured in $HNO_3$. The same preparation but using palygorskite from Attapulgus instead of that from Mexico showed a light blue colour after reflux. Pigments made with natural leaves (one made by Reyes-Valerio in 1990 and another produced for this work) were strongly decoloured, but a light blue tonality remained. Pigments containing more synthetic indigo (3% and 7%) were always blue after reflux in $HNO_3$, but the intensity of the blue decreased. XRD on samples after reflux shows significant destruction of palygorskite with the formation of amorphous silica in all cases (Fig. 5). More intensive attacks were attempted in order to fully destroy the pigment. $HNO_3$ reflux at 90 ºC during 30 h did not completely destroy the palygorskite clay and XRD peaks from the palygorskite are still present (Fig. 5). Moreover, it presented a blue tonality. A reflux with HCl at boiling point, during 30 hours completely destroyed the pigment, as observed by XRD, and was accompanied by a complete decolouration, resulting in a white powder.

## *4.5 Discussion*

When comparing pigments made with palygorskite to those made with sepiolite, it has been found that a fairly stable pigment (i.e., resistant to few minutes in acid) is easily obtained using sepiolite, but it is not stable to long attacks (more than 1.5 h). The sepiolite pigments stabilize at lower temperatures (starting from room temperatures, where the clay grain-size seems to be a relevant parameter) than the palygorskite pigments. However, all sepiolite pigments are destroyed in acids at room temperature. Palygorskite pigments were not affected by acids at room temperature in both colouration and crystallinity.

The acid attacks affect more sepiolite than palygorskite because of two reasons: i) the higher contents of $Mg^{2+}$ ions in octahedral environment (Suárez Barrios et al. 1995), that are primarily affected by the acid, and ii) the larger size of the channels of sepiolite (5.6 × 11 Å) than those of palygorskite (3.7 × 6 Å) thus making the internal surfaces more accessible to aggressive agents. When the acid dissolves the Al and $Mg^{2+}$ octahedra, the spatial arrangement of the Si atoms, surrounded each one by a tetrahedron of oxygens, is altered (the tetrahedra do not change, but the arrangement is destroyed), therefore loosing the crystallinity. Amorphous silica is produced, presenting, however, the morphological aspect (fibrous structure) of its precursor, as it can be observed by TEM (Suárez Barrios et al. 1995).

Considering that only few studies of synthetic Maya blue with sepiolite are reported ((Hubbard et al. 2003; Van Olphen 1966), which do not detail the acid tests performed), contrary to the numerous studies done with palygorskite, our results confirm an original behaviour of sepiolite - pigments to acid aggressions, in two senses. On one hand, a pigment fairly resistant to $HNO_3$ during several minutes can be obtained without heating (not true for palygorskites). On the other hand, the sepiolite pigments are not resistant (not only decoloured, but its clay structure is destroyed) to concentrated acids at room temperature for a long time. The sepiolite-based pigments, although more resistant than those made with planar clays, like montmorillonite, are not resistant to long exposures to concentrated acids towards to a total destruction of the sepiolte.

The interest of the study of sepiolite pigments is more in the chemical context rather than in the archaeological one. The differences observed between sepiolite and palygorskite in terms of acid attack, thermal behaviour, chemical composition, etc. could be crucial for unambiguously determine the characteristics of the chemical bonding between the indigo and the clay in Maya blue. From the archaeological point of view it is not so interesting, as almost all archaeological samples studied in literature show a major proportion of palygorskite. Sepiolite, when found, could be only at the level of impurity together with other clays and minerals. If we consider that one of the characteristics of the Maya blue is its high resistance to aggressive acid treatments (as it seems to be the present consensus), sepiolite should be excluded as a clay material for the Maya blue.

All palygorskite pigments, as well as raw palygorskite are resistant to concentrated acids at room temperature. Treatments of several days do not change at all the XRD nor decolour the pigments. It is necessary to treat the pigments in hot (90 ºC) acids during 5 hours to observe a partial decolouration and a significant destruction of the palygorskite. Total destruction of the pigment and raw palygorskite is possible after 30 h treatment in 6 mol.l$^{-1}$ HCl in ebullition.

## 5 Summary and conclusions

We have synthesized several Maya blue samples using different methods proposed in literature: the "laboratory" methods proposed by van Olphen (1966) and the "traditional" method proposed by Reyes-Valerio (1993), probably used by the Mayas. The latter method mixes up leaves from the *Indigofera suffruticosa* plant, and palygorskite clay (known by the local Maya people as *sak lu'um,* literally "white-earth"). All preparation methods used in this work have in common a heating phase after the ingredients have been combined and dried. The simplest way to prepare the pigment is to mix palygorskite clay with indigo powder (about 1%), grind the mixture in a mortar, and heat to 150-200 ºC for more than one hour.

Our results on acid-resistance tests allow to draw the following conclusions:

- the indigo does not protect the clay against acid-attacks, because similar destructions are observed in the case of pigments and raw clays. The resistance to the acid is not due to the role of the indigo, but to the role of the clay: the destruction of the clay implies the destruction of the pigment.

- the palygorskite and sepiolite clays react with indigo producing a pigment which does not decolour in $HNO_3$, as it happens for indigo. This is not true for the laminar clays (montmorillonite). It shows that the fibrous structure of palygorskite and sepiolite featuring channels is important for the stabilization.

- raw sepiolite and sepiolite pigments are much more affected but the acids than palygorskite and palygorskite pigments. Sepiolite pigments are destroyed with long treatments (from several hours to a few days) in concentrated acids at room temperature.

- palygorskite pigments are very resistant: no changes were observed neither in colour nor in crystallinity after long acid attacks at room temperature. Decolouration and destruction of the pigment matrix are observed with reflux treatments at 90 ºC during 5 h, and complete destruction is achieved with very strong attacks (6 N HCl in ebullition, 30 h).

- The decolouration it is not always strongly correlated to destruction of the clay matrix. In some cases a complete decolouration is observed (in sepiolite pigments) and not all the clay is destroyed. On the other hand, we sometimes obtained XRD profiles with almost no peaks, and the residue of the attack presented a light blue colour. However, total destruction of the palygorskite pigment was accompanied by total decolouration. The HCl tends to destroy the clay more intensively than $HNO_3$, but, on the contrary, $HNO_3$ decolours more the pigments than HCl (because, perhaps, of its evident effect on indigo).

The results presented here clearly show that the high resistance of Maya blue to acid attacks is certainly due to the exceptional resistance of the palygorskite clay. The description of the chemical bond linking indigo and palygorskite is still the key problem that will certainly need further studies. Whether the indigo penetrates or not into the palygorskite channels is not the main question now, because indigo does not make the Maya blue more acid-resistant than the raw palygorskite. New investigations should also focus on the nature of indigo molecule, how it links with some atomic groups of the palygorskite, why is moderate heating needed, what are the chemical reasons of the characteristic Maya blue colour, and why is the colour of the new indigoid complex present in Maya blue is not affected by nitric acid as indigo is.

## Acknowledgements

Thanks to J. Santarén and the Spanish company TOLSA for generously sending us samples of palygorskite and sepiolite and for helpful information concerning these clays. N. Peltier collaborated in preparing most of the pigments. M. Anne is acknowledged for his continuous support, and L. Alianelli and R. Felici for helpful discussions. Also thanks to P. Strobel for his support and advise in Chemistry and to P. Bordet for his help with the XRD diffractometer. Thanks also to S. Tremeau for her help in the chemistry laboratory.